\documentclass[noshowpacs,amsmath,twocolumn,superscriptaddress,10pt,aps,notitlepage,nofootinbib]{revtex4-1} 
\usepackage{setspace}
\usepackage{amsmath}
\usepackage{breqn}
\usepackage{graphicx}
\usepackage[nearskip,margin = 0pt,caption=false]{subfig}

\usepackage{verbatim}
\usepackage{amsfonts}
\usepackage{amssymb}
\usepackage{epstopdf}
\usepackage{lipsum}
\usepackage{siunitx}
\usepackage{xcolor}
\usepackage{array}
\usepackage[normalem]{ulem}
\usepackage[resetlabels,labeled]{multibib}
\DeclareGraphicsExtensions{.pdf,.eps,.png,.jpg,.mps}
\usepackage{etoolbox}

\begin{document}
\title{Inverse-Designed Photonic Crystal Circuits for Optical Beam Steering}
\author{Dries Vercruysse$^{1}$, Neil V. Sapra$^{1}$, Ki Youl Yang$^{1}$, and Jelena Vu\v{c}kovi\'{c}$^{1}$\\
\vspace{+0.05 in}
$^1$E. L. Ginzton Laboratory, Stanford University, Stanford, CA, USA.}

\begin{abstract}
The ability of photonic crystal waveguides (PCWs) to confine and slow down light makes them an ideal component to enhance the performance of various photonic devices, such as optical modulators or sensors.
However, the integration of PCWs in photonic applications poses design challenges, most notably, engineering the PCW mode dispersion and creating efficient coupling devices.
Here, we solve these challenges with photonic inverse design, and experimentally demonstrate a slow-light PCW optical phased array (OPA) with a wide steering range.
Even and odd mode PCWs are engineered for a group index of 25, over a bandwidth of 20nm and 12nm, respectively.
Additionally, for both PCW designs, we create strip waveguide couplers and free-space vertical couplers.
Finally, also relying on inverse design, the radiative losses of the PCW are engineered, allowing us to construct OPAs with a 20$^o$ steering range in a 20nm bandwidth.
\end{abstract} 

\maketitle


\begin{figure*}[t!]
\centering
\includegraphics[width=0.95\linewidth]{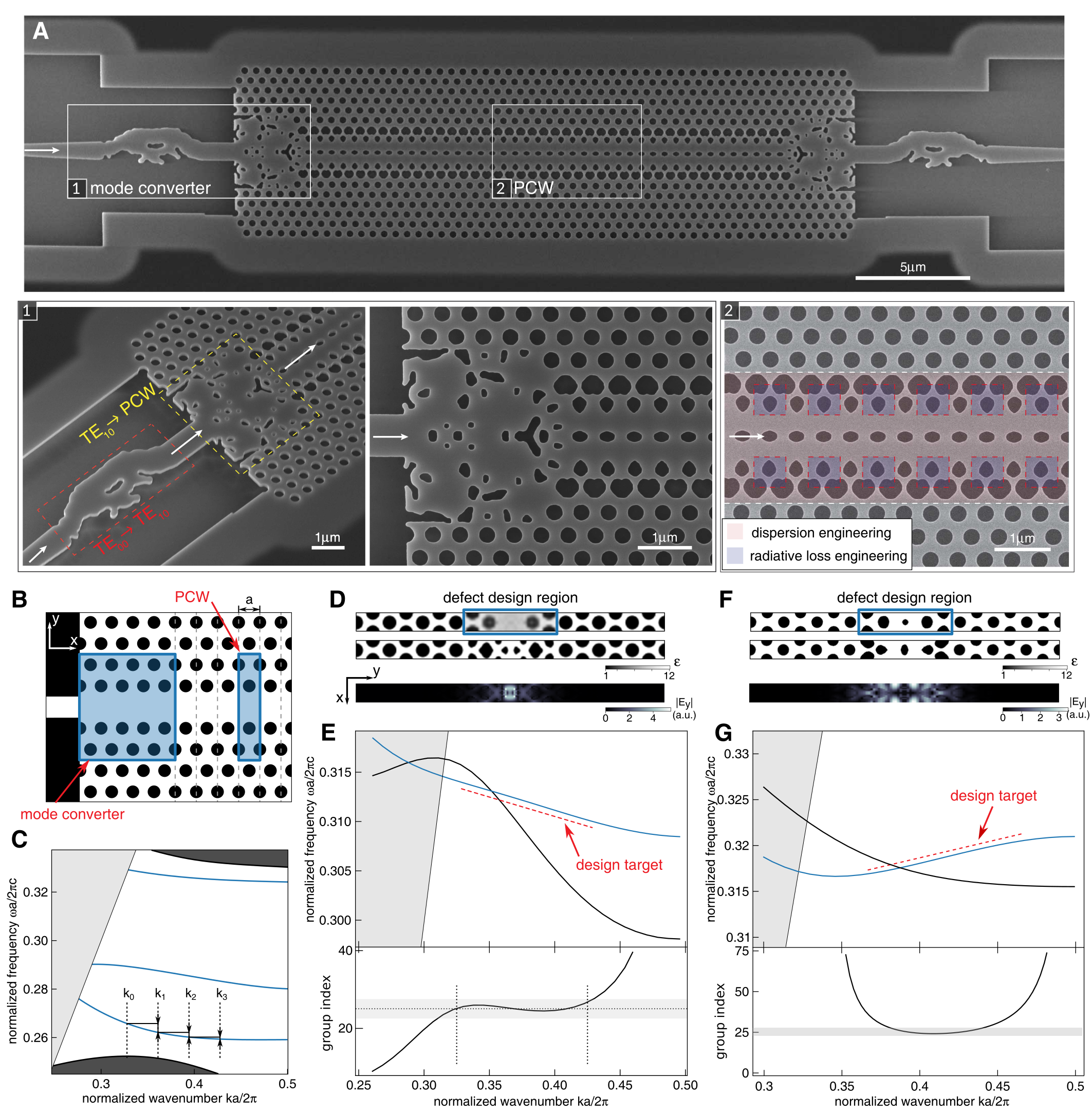}
\caption{\label{fig:InvDes} 
\textbf{Dispersion engineering with inverse design.} 
(\textbf{A}) SEM micrograph of an inverse designed slow-light PCW device. 
Insets: a detailed image of a two-stage mode converter is shown in panel 1. 
Panel 2 shows the inverse designed PCW with the design regions for dispersion engineering in red and radiative loss engineering in blue. 
(\textbf{B}) Design regions in a slow-light device. 
The right blue area depicts the design region of the photonic crystal waveguide with period $a$; the left blue region indicates the design region for the strip-to-PCW mode converter. 
(\textbf{C}) Band diagram of W1 photonic crystal defect line. On the bottom guided mode, the frequency difference for four wavenumbers are indicated.
(\textbf{D, F}) Initial (top), and final (middle) permittivity and $|E_y|$ field profiles (bottom) for an even mode (\textbf{D}) and odd mode (\textbf{F}) PCW. 
(\textbf{E, G}) Top: dispersion relation of the initial (black) and final (blue) even mode slow-light PCW inverse design optimization (\textbf{E}). The odd mode slow-light PCW optimized dispersion relation is shown in (\textbf{G}). 
The dashed line indicates the target slope. 
The gray area represents the region above the light-line.
Bottom: group index of the final structure. The gray area indicates the $\pm$ 10\% deviation around the center group index.
}
\end{figure*}

\begin{figure*}[t!]
\centering
\includegraphics[width=0.95\linewidth]{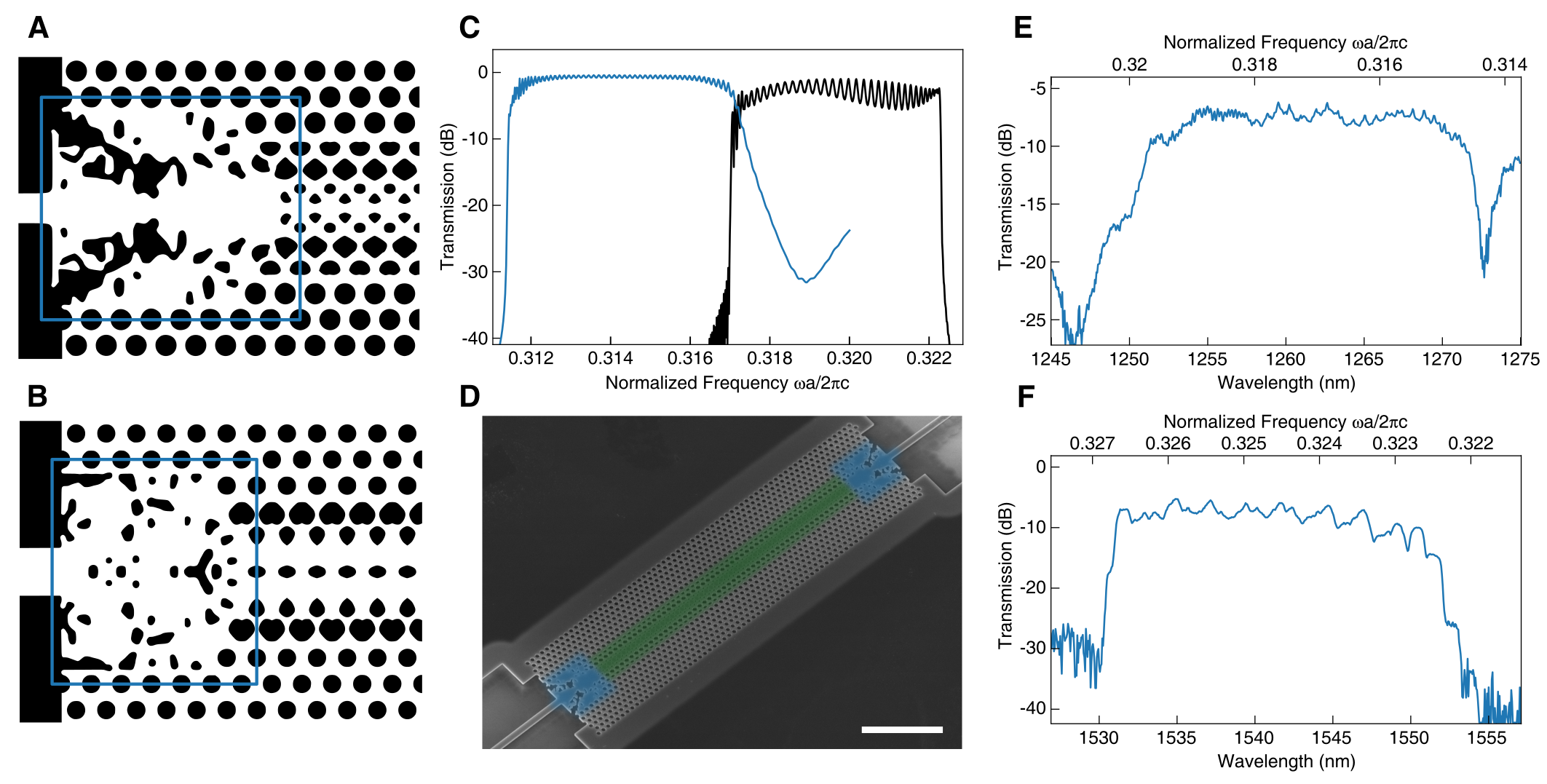}
\caption{\label{fig:coupler} 
\textbf{Strip-to-PCW mode converters.} 
(\textbf{A, B}) Final optimized strip-to-PCW converters for the even and odd mode slow-light PCWs, respectively. 
The blue rectangle indicates the design region. 
(\textbf{C}) Simulated transmission spectra for a \SI{20}{\micro\meter} long delay line with  even (blue) and odd (black) PCWs. 
(\textbf{D}) Micrograph of a \SI{20}{\micro\meter} long delay line (bar: \SI{5}{\micro\meter}). 
(\textbf{E, F}) Experimentally measured transmission spectra through a \SI{60}{\micro\meter} long delay system based on an even and odd slow-light PCW, respectively.
}
\end{figure*}

\begin{figure*}[t!]
\centering
\includegraphics[width=0.95\linewidth]{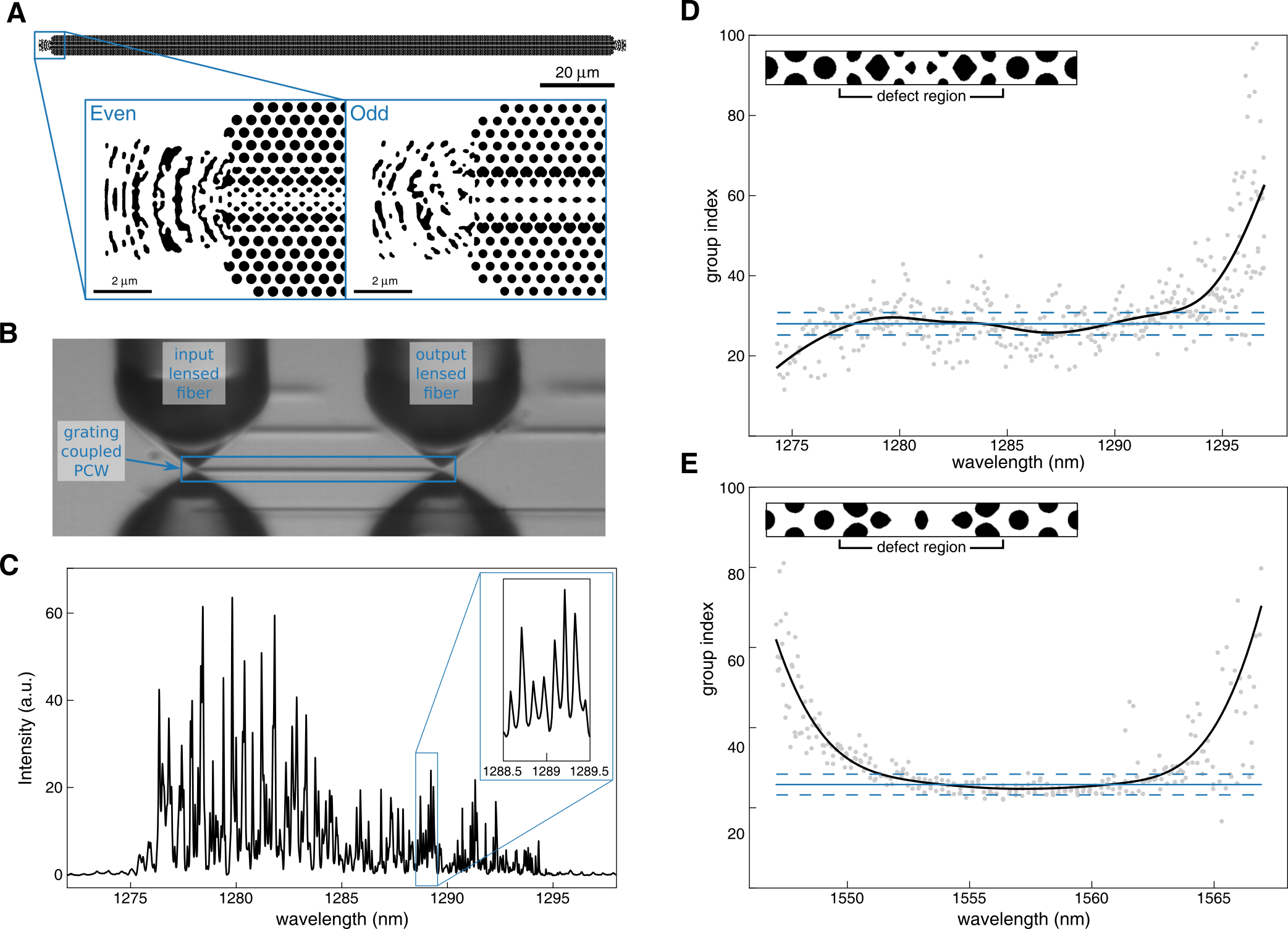}
\caption{\label{fig:DispMeas}
\textbf{Dispersion measurement.}
(\textbf{A}) Slow light waveguide with free-space vertical couplers. 
The insets show the inverse designed free-space couplers for the even and odd mode. 
(\textbf{B}) Microscope image of the measurement setup using lensed fibers for vertical coupling. 
(\textbf{C}) Transmission spectrum of an even mode slow-light PCW through the inverse designed vertical free-space couplers. 
(\textbf{D, E}) Group index calculated from the free spectral range of transmission spectra through an even (\textbf{D}) and odd (\textbf{E}) mode slow-light PCW.
The figures combine the results of three measured devices. 
The black curve is a fit to the grey measurement points. 
The solid blue line indicates the average group index over the operating wavelength interval. 
The dashed blue lines indicate a 10\% deviation from the average.
}
\end{figure*}

\begin{figure*}[t!]
\centering
\includegraphics[width=0.95\linewidth]{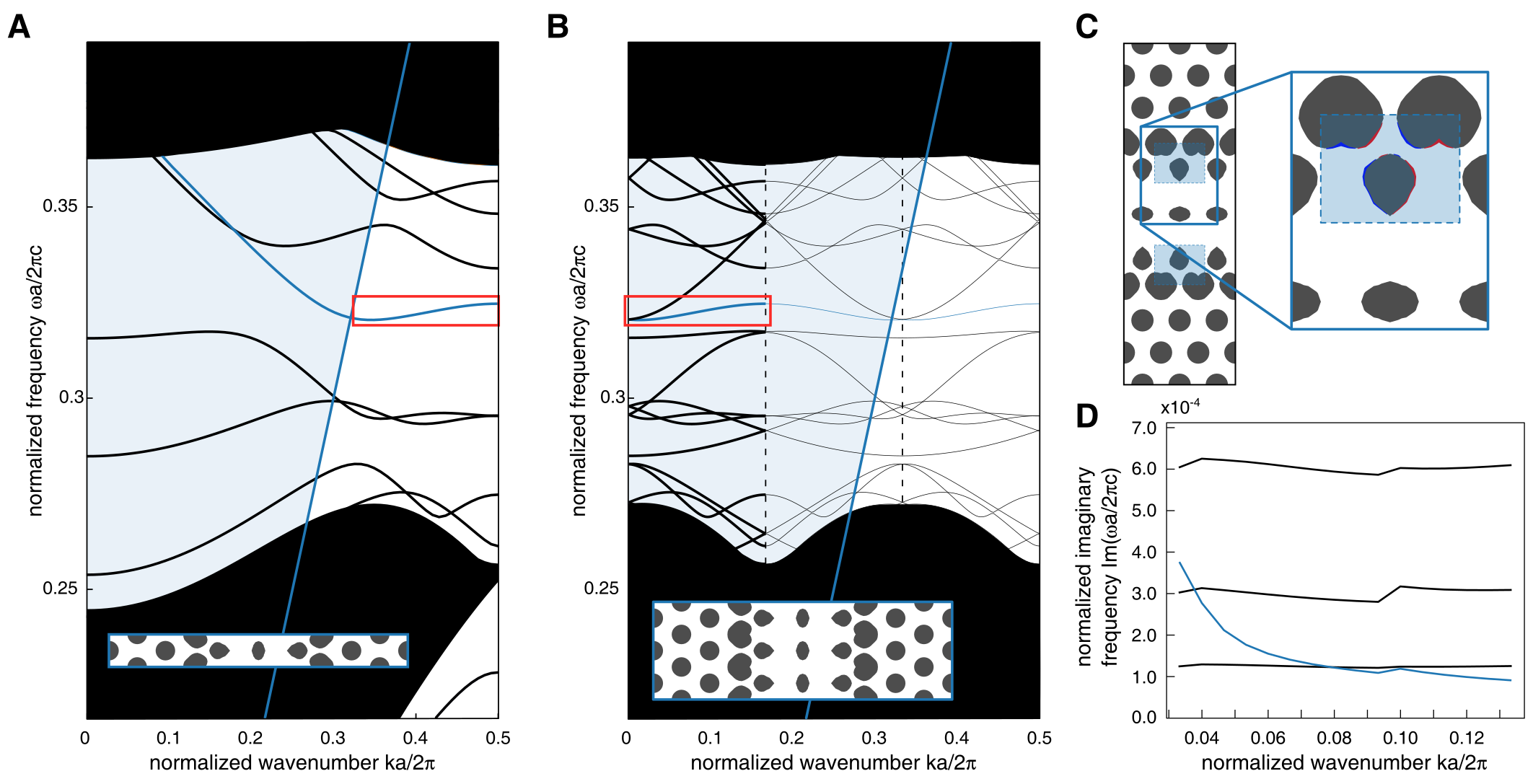}
\caption{\label{fig:losses} 
\textbf{Radiative loss engineering.} 
(\textbf{A}) Full band diagram for an odd mode PCW. Slow-light region is indicated in red rectangle.  
(\textbf{B}) Band diagram for the modified radiative PCW, with period expended to be three times that of the non-radiative odd mode PCW. Slow-light region is indicated in red rectangle.  
(\textbf{C}) Zoom-in of modified radiative PCW: removed material is indicated in red, and added material in blue.
(\textbf{D}) Normalized imaginary frequency spectrum for randomly perturbed photonic crystal structure in blue, and optimized spectra in black for a target normalized imaginary frequency of \SI{1.25e-4}, \SI{3.0e-4} and \SI{6.0e-4}.
}
\end{figure*}

\begin{figure*}[t!]
\centering
\includegraphics[width=0.95\linewidth]{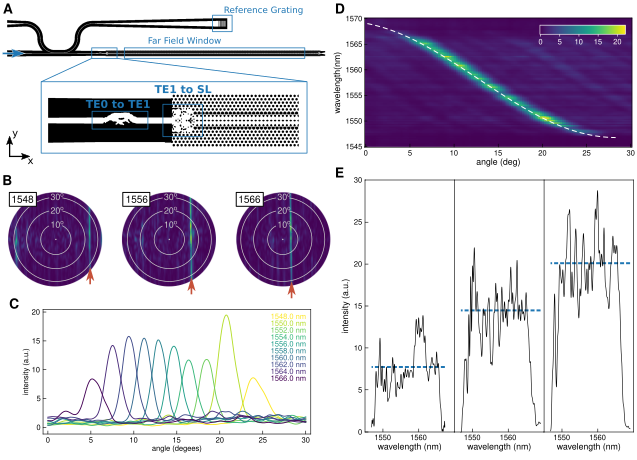}
\caption{\label{fig:OPA} 
\textbf{Inverse designed slow-light OPA.} 
(\textbf{A}) Measurement circuit for the OPA. A directional coupler is used to route light to a reference grating for power monitoring. Transmitted light is sent to the slow-light OPA through the use of inverse designed mode converters (inset). 
(\textbf{B}) Back focal plane measurements of the far-field window indicated in (\textbf{A}) for three wavelengths, demonstrating beam steering. 
(\textbf{C}) Angular spectrum for different wavelengths. 
(\textbf{D}) OPA radiation intensity heatmap as a function of angle and wavelength. The dashed white line, fit to the peaks, indicates where the intensity is sampled for the spectra in panel E. 
(\textbf{E}) Intensity spectra for three OPAs designed with different target imaginary frequencies (radiative losses). The blue line indicates the average radiation intensity over the operating bandwidth.
}
\end{figure*}

\section*{Introduction}
Photonic crystal waveguides (PCWs), constructed by a defect line in a photonic crystal lattice, possess remarkably tuneable optical modes.
The properties of these modes can be engineered for various photonic applications, such as nonlinear photonic devices or emission systems that can achieve increased efficiency and reduced device size. 
This is accomplished by relying on the ability of the PCW to spatially compress optical energy and enhance light-matter interactions
\cite{
Soljacic2004, 
Corcoran2009a,
Javadi2015, 
Marty2020,
Zhou2014,
Ek2014a}.
Additionally, modulators and sensors use slow-light PCWs for their high phase-velocity sensitivity to the refractive index of the waveguide or surrounding medium
\cite{
Nguyen2010, 
Baba2014, 
Hinakura2019, 
Yan2017a, 
Nozaki2010,
Qin2016,
Zhao2012,
Wang2010}.
Recently, slow-light PCWs were shown to be an effective foundation for optical phased arrays (OPAs), providing improvements in their beam steering capabilities
\cite{
Takeuchi2018,
Abe2018,
Ito2018,
Ito2020}.
The wavelength-dependent steering sensitivity of the OPA depends on the dispersion relation of the waveguides that feed the OPA antennas. 
The smaller the group velocity of the waveguides, the larger the beam deflection by a change in wavelength 
\cite{
Dostart2020,
Costantini2015,
Poulton2019,
Raval2017,
Sun2013,
Miller2020,
Hulme2015,
Acoleyen2011}.
Therefore, a slow-light PCW can be used to increase the OPA steering range while reducing its operating bandwidth.

PCW design is complicated, presenting a hurdle to its widespread adoption despite the many benefits.
The challenge is twofold: first, the photonic crystal defect line needs to be modified to match the desired dispersion relation of the application, and second, converters are required to efficiently couple to the PCW mode.
Traditional dispersion engineering methods typically vary a few geometric parameters, such as the position or radius of lattice holes around the defect line
\cite{
Zhao2015,
Serna2016,
Rahimi2011a}.
In contrast, inverse design methods can optimize over any number of degrees-of-freedom, resulting in designs with arbitrary shapes
\cite{Molesky2018,
Abe2018}.
These techniques have produced compact and efficient devices, making inverse design a promising solution for the dispersion engineering of PCWs
\cite{Burger2004,
Preble2005,
Riishede2008,
Wang2011,
Matzen2011,
Vercruysse2020}.
Additionally, poor modal and group-index mismatch results in high PCW insertion loss, which needs to be mitigated through mode converters.
Traditionally, for an unaltered defect line, the lattice can simply be tapered; however, this approach is not applicable for a defect line with modifications
\cite{Rahimi2011a}.
Inverse design steps away from this limited approach and explores the entire design space, allowing for PCW coupling solutions otherwise inaccessible.
\cite{Matzen2011,
Vercruysse2020}.

In this work, we demonstrate the use of inverse design for PCW-based photonic systems. 
Here, we construct a compact, slow-light OPA with a wide steering angle and small operating bandwidth.
First, we illustrate how inverse design can be used for dispersion engineering by constructing  slow-light PCWs for both the even and odd optical mode.
For both modes, coupling solutions to the slow-light PCW are provided in the form of strip waveguide converters, as well as free-space vertical couplers.
We then measure the coupling efficiency, dispersion, and waveguide losses of the slow-light devices (Fig. \ref{fig:InvDes}A). 
Finally, we modify the slow-light PCW to intentionally introduce radiative losses (Fig.\ref{fig:InvDes}A.2), also through inverse design. 
We show that the radiation can be optimized to a target value while keeping the dispersion relation unaltered.
Three OPA designs, each with a different radiative strength, are experimentally demonstrated to sweep \ang{20} of angular range over \SI{20}{\nano\meter} of optical bandwidth.
These results mark a stark contrast to OPAs based on silicon strip waveguides, which sweep about \ang{15} in \SI{100}{\nano\meter} optical bandwidth 
\cite{Acoleyen2011}.\\


\section*{Results}


\noindent\textbf{Inverse design dispersion engineering.}  
Slow-light devices require a high group index which stays constant over an operating bandwidth; this requires a linear dispersion relation with a slope that matches the target group index.
In designing our slow-light PCW, we begin with a W1 defect line PCW.
The dispersion relation of the W1 mode is highly nonlinear, and so dispersion engineering is required for slow-light operation.
We optimize the PCW for linear dispersion by altering a region around the line defect, which encompasses two rows on either side of the waveguide (right box in Fig. \ref{fig:InvDes}B).
To demonstrate the flexibility of this method, two types of slow-light PCWs are targeted: an even mode design with a negative-slope dispersion, and an odd mode design with a positive-slope dispersion.
For both the even and odd PCWs, the dispersion was optimized for a target group index of 25.
To achieve this, our inverse design method aims to equalize the pair-wise frequency differences for a series of wavenumbers, $k_i$, to a target difference that matches the desired group index.
This procedure is illustrated in Fig. \ref{fig:InvDes}C for our W1 line defect, using four wavenumber values.

The two slow-light PCWs are designed for a \SI{220}{\nano\meter} thick silicon device layer, with air above and below. 
The initial structure for the even mode PCW is shown in the top panel of Fig. \ref{fig:InvDes}D.
This unit cell, derived from a W1 defect line, has a center design region with permittivity values between that of the waveguide and background cladding.
The operating wavelength of the design is targeted to be \SI{1300}{\nano\meter};
therefore, to target the defect line modes, we choose the lattice constant of the photonic crystal, $a$, to be \SI{400}{\nano\meter}.
The dispersion relation of the initial structure is shown in Fig. \ref{fig:InvDes}E (black curve). 
After optimization, the final structure (middle Fig. \ref{fig:InvDes}D) has a dispersion that is linear in the $k \in [0.325\frac{2\pi}{a}, 0.425\frac{2\pi}{a}]$ design interval (blue curve Fig. \ref{fig:InvDes}D).
The group index of the PCW based on this curve is shown in the bottom graph of Fig. \ref{fig:InvDes}E. 
The average group index within the design interval is 25.14. 
Considering a $10\%$ margin of error (grey area in bottom graph, Fig. \ref{fig:InvDes}E), the wavenumber interval can be extended to $ [0.31\frac{2\pi}{a}, 0.43\frac{2\pi}{a}]$, resulting in a normalized bandwidth group index product, NBGP, of $0.38$ 
\cite{Zhai2011}.

The odd mode is optimized in a similar fashion. 
However, here, we choose to start from the defect line shown in Fig \ref{fig:InvDes}F. 
The design was aimed for an operating wavelength of \SI{1550}{\nano\meter}, and thus a larger lattice constant of \SI{500}{\nano\meter} is required. 
The final optimized structure (middle graph of Fig. \ref{fig:InvDes}F) has a dispersion relation (Fig. \ref{fig:InvDes}G) which shows a positive  slope matching the target group index of 25. 
Allowing for $10\%$ deviation, the waveguide has an operating $k$-interval of $[0.38\frac{2\pi}{a}, 0.45\frac{2\pi}{a}]$, a center group index of 26.8, and an NBGP of 0.22.\\  


\noindent\textbf{Slow-light system design and measurements.}  
Coupling to photonic crystal waveguides from a conventional strip or rib waveguide often results in low efficiency due to modal and group index mismatch.
While this can be mitigated by tapering, these solutions are particularly difficult for photonic crystal waveguides that deviate from simple line defects.
For both PCW designs, we created mode couplers using inverse design techniques
\cite{Vercruysse2020, Vercruysse2019}.
A design region, consisting of several periods at the start of the PCW, is allocated for the mode converter (left box in Fig. \ref{fig:InvDes}B).
The inverse design optimization maximizes the PCW transmission over the slow-light operating bandwidth.
Fig. \ref{fig:coupler}A shows the final design for the even-mode slow-light PCW converter, operating over the frequency interval, $\frac{\omega a}{2\pi c} = [0.309, 0.314]$.
The device couples the fundamental TE0 mode of a \SI{400}{\nano\meter} width strip waveguide to the even slow-light mode.
Similarly, we designed a coupler for the odd slow-light mode (Fig. \ref{fig:coupler}B), with an operating frequency interval $\frac{\omega a}{2\pi c} = [0.322, 0.326]$.
Since the targeted slow-light PCW mode is odd, the input strip waveguide mode is likewise chosen to be the odd TE1 mode.
A coupler from the TE0 strip waveguide mode to the odd PCW mode can also be generated through inverse design (Suppl. Info. S2); however, here we opted for a combination of a TE0-to-TE1 converter, followed by a TE1-to-PCW converter, to produce a symmetric design.

The performance of these converters is validated through FDTD simulations in the configuration of photonic delay lines.
Two identical converters are reflected back-to-back, separated by \SI{20}{\micro\meter} of their respective photonic crystal waveguide.
Simulated transmission spectra for the even and odd mode can be seen in Fig \ref{fig:coupler}C.
A clear transmission window over the frequency interval targets for both converters is achieved.

In order to experimentally characterize our devices, we fabricate slow-light waveguide delay lines of varying lengths on \SI{220}{\nano\meter} thick silicon-on-insulator, using electron beam lithography (see methods).
An SEM micrograph of a \SI{20}{\micro\meter} long suspended device can be seen in Fig. \ref{fig:coupler}D.
The transmission spectrum of a \SI{40}{\micro\meter} long, even mode device (Fig. \ref{fig:coupler}E) shows a transmission window around \SI{1252}-\SI{1272}{\nano\meter}, with a relatively flat transmission of around \SI{-7.6}{\decibel} (average over interval).
The different delay line lengths allow us to characterize the insertion and waveguide losses as well.
For the even PCW mode, the transmission analysis shows \SI{-22}{\decibel\per\milli\meter} loss and an average \SI{-3.4}{\decibel} insertion loss for the coupler (Suppl. Info. S3).
Similar characterization for the odd mode showed a waveguide loss of \SI{-10}{\decibel\per\milli\meter} and a \SI{-4.5}{\decibel} average insertion loss for the coupler.
A transmission spectrum for a \SI{40}{\micro\meter} long delay line can be seen in Fig. \ref{fig:coupler}F and shows a transmission window at \SI{1532}-\SI{1552}{\nano\meter}.
Recall, the odd mode device relied on an additional TE0-to-TE1 mode converter as seen in Fig. \ref{fig:InvDes}A. 
Additional information on the TE0-to-TE1 mode converter can be found in the supplementary material (Suppl. Info. S1).

The significantly lower losses for the odd mode are to be expected from the designs (Fig. \ref{fig:InvDes}D) and simulated field profiles (Fig. \ref{fig:InvDes}F). 
The even mode design has multiple small features at the center of the defect line, where the field intensity is largest.
Fabrication errors and surface etch roughness, which are the primary loss causes, would therefore impact the even mode design more than the odd mode.\\


\noindent\textbf{Dispersion measurements.}  
The dispersion of a waveguide can be derived from the free spectral range (FSR) of a Fabry-Perot resonator.
For slow-light waveguides, these resonators are typically realized by edge terminating the waveguide on each side.
However, edge termination suffers from low coupling efficiency, especially for a high group index.
Moreover, to measure the dispersion of the odd mode, a higher order Gaussian beam would need to be used, further convoluting the measuring setup. 
Instead of edge couplers, here we rely on inverse designed vertical couplers, which couple from free-space directly to the PCW mode and form a low-Q Fabry-Perot resonator.
Fig. \ref{fig:DispMeas}A shows a \SI{200}{\micro\meter} Fabry-Perot resonator and the inverse designed vertical couplers for the even and odd mode.
For the odd mode, the input Gaussian and output mode break symmetry, which results in the asymmetric design.
More information on the vertical couplers can be found in \cite{Vercruysse2020}.

The transmission measurement setup uses lensed fibers with a \SI{2}{\micro\meter} spot diameter to couple to the in- and output couplers (Fig. \ref{fig:DispMeas}B).
A transmission spectrum for an even mode Fabry-Perot resonator can be seen in Fig. \ref{fig:DispMeas}C and shows a dense fringe pattern.
The group index is derived from the local FSR of the transmission spectra, and is shown in Fig. \ref{fig:DispMeas}D and E for both even and odd modes, respectively.
A univaried spline fit through the points (black curve) matches the simulated results of Fig. \ref{fig:InvDes}E and G.
Based on the fit, the even mode has an average group index of 27.98 over a \SI{1277}{\nano\meter} to \SI{1293}{\nano\meter} spectral window ($\pm10\%$-error), resulting in an NBGP of 0.34.
The odd mode waveguide has an NBGP of 0.19 with an average group index of 25.83 in a \SI{1552}{\nano\meter} to \SI{1563}{\nano\meter} spectral window.


\noindent\textbf{Radiative loss engineering.}  
By introducing periodic perturbations to a PCW, light can be directionally radiated from the PCW.
In concert with the flat dispersion curve of the slow-light PCW, a small change in wavelength translates to a large change in the direction of the radiated beam.
This provides the basis for a wavelength-tunable, large-angle steering optical phased array (OPA).

The slow-light dispersion relation is sensitive to small alterations in the structure.
As such, introducing a periodic perturbation that does not alter the dispersion is challenging.
To overcome this, we rely on inverse design to engineer radiative loss in the slow-light waveguide, while still maintaining a flat dispersion relation.
The objective function for this inverse design optimization problem consists of two terms.
The first term aims to push the imaginary part of the frequency, which represents the radiative loss, to a target value.
The second term penalizes any deviation from the original dispersion relation. 
The objective thus relies on both the real and imaginary components of a band diagram calculation.
For the band diagram calculations, we use the guided mode expansion software, Legume, which we incorporated into the SPINS inverse design framework \cite{Minkov2020, Su2019}.

The radiation pattern of an OPA ideally consists of a single lobe. 
Light scattered from a defect in the even mode slow-light PCW was observed to radiate in a quadrupole pattern, producing multiple lobes in the far-field (Suppl. Info. S4).
In contrast, the light scattered from the odd mode radiates in a dipole pattern, producing a single spot in the far-field.
Due to its desirable properties, we choose the odd mode slow-light PCW as the starting point for our radiation optimization.

The full band diagram of the odd mode is shown in Fig. \ref{fig:losses}A.
As is typical for slow-light waveguides, the optimized band region (red rectangle) is located under the light-line, as to avoid radiative losses. 
For the OPA, we want to introduce radiative losses, and so we want to move the band above the light-line.
This can be done by breaking the original periodicity and considering a larger period.
We choose to work with a new unit cell that is three times the period of the original unit cell (inset Fig. \ref{fig:losses}B), and make small modifications to this larger unit cell.
The band diagram of the three-period unit cell, which resembles the original band diagram folded in three, is now located above the light-line (Fig. \ref{fig:losses}B).
The radiative losses of the mode will depend on the extent of the modifications.
Since OPAs are typically millimeters to centimeters in length, we aimed for low radiative losses.
The modifications required will then necessarily be small perturbations, which may pose a problem for fabrication.
We therefore limit the design region to be the first two rows of the defect line (Fig. \ref{fig:losses}C), where the fields are of smaller magnitude, compared to the center of the waveguide. 
This will result in larger modifications to achieve the target losses.

The radiative losses of the three optimized devices with target normalized imaginary frequencies of \SI{1.25e-4}, \SI{3.0e-4}, and \SI{6.0e-4}, are shown in Fig. \ref{fig:losses}D.
The losses for a waveguide with random alterations, as used as an initial condition, are shown in blue.
After optimization, the imaginary frequency over the k-space is flat around the target value.\\


\noindent\textbf{Beam Steering.}  
The inverse designed loss- and dispersion-engineered PCWs, in combination with the couplers, allow us to realize a slow-light optical phased array.
The OPA photonic test circuit is shown in Fig. \ref{fig:OPA}A.
Light is coupled to the bottom-left waveguide via an edge coupling setup, and a directional coupler is used to route a small amount of power to a reference grating coupler.
The signal from this grating coupler is used as a reference to normalize the power from the OPA.
The transmitted light from the directional coupler passes through a TE0-to-TE1 mode converter, and subsequently goes to the inverse designed OPA.
The OPA is constructed with the TE1 strip-to-PCW converter, followed by twenty periods of the non-radiative odd mode slow-light PCW, after which the PCW transitions to the radiative design.
To minimize reflections, the waveguide is terminated through a transition to a tapered strip waveguide, as facilitated by the  strip-to-PCW converter.
We evaluate the direction and intensity of the radiated OPA light with a back-focal-plane (BFP) microscope (Suppl. Info. S6).
A rectangular aperture in the bright-field image plane of the BFP microscope is used to filter out all but the light emitted by the OPA-waveguide (far-field window in Fig. \ref{fig:OPA}A).
BFP images at three different wavelengths are shown in Fig. \ref{fig:OPA}B.
Since the OPA consists of a single waveguide, the radiated light is only confined in the x-direction, and not in the y-direction.
The BFP images thus show a line, rather than a point.
As the wavelength is swept from \SI{1548}{\nano\meter} to \SI{1566}{\nano\meter}, the radiated light is steered from an angle of $\theta$\textsubscript{x} = \ang{24} to $\theta$\textsubscript{x} = \ang{5}.
The angular intensity spectra can be obtained by integrating the BFP intensity along the $\theta$\textsubscript{y} direction (Fig. \ref{fig:OPA}C).
The heat map constructed from these spectra shows the radiation intensity as a function of $\theta$\textsubscript{x} and the wavelength (Fig. \ref{fig:OPA}D).
The dotted line fits the maxima of the angular spectra; as expected, this curve corresponds to the dispersion relation shown in Fig. \ref{fig:losses}B.
The intensity is sampled along the fitted steering curve, and is shown in the middle panel of Fig. \ref{fig:OPA}E.
On the left and right panels, the same analysis is shown for OPAs designed with the target imaginary frequency values (radiative loss) of \SI{1.25e-4} and \SI{6e-4}, respectively.
The average intensity, indicated by a blue dashed line, follows the trend of engineered radiative loss, as shown in Fig. \ref{fig:losses}D.
The radiation intensity as a function of wavelength and angle for all three OPAs can be found in the supplementary information (Suppl. Info. 7S). \\


\section*{Conclusion}
We have demonstrated the use of inverse design methods to tackle many of the challenges associated with photonic crystal circuits -- specifically in the context of on-chip beam steering.
Inverse design was shown to be an effective tool for dispersion engineering the modes of a PCW, as well as for developing multiple coupling solutions to these modes.
These devices allowed us to create PCW delay lines and resonators, which we used to verify the group index spectra and characterize the losses experimentally.
Furthermore, we showed that radiative losses can be added to the dispersion optimization procedure.
Relying on these loss- and dispersion-engineered waveguides, we constructed three OPAs with different tailored radiative strengths.
A steering range of \ang{20} over a \SI{1548}-\SI{1566}{\nano\meter} wavelength interval was experimentally verified. 
These inverse designed slow-light OPAs represent a significant improvement in both steering range and operating bandwidth, as compared to conventional OPAs based on silicon strip waveguides, which steer around \ang{15} over a \SI{100}{\nano\meter} bandwidth.
Moreover, inverse design was shown to be highly effective in tackling the multitude of design challenges in these photonic systems.
Through this work, we illustrated the potential of this method to enable innovative photonic crystal architectures, which can find applications in nonlinear optics, Mach-Zehnder modulators, topological photonics, and dispersion engineering for microcombs or quantum optics.\\

\vspace{1cm}

\noindent{\bf Methods}\\
\\
\noindent\textbf{Inverse design} All inverse design optimizations were performed with the Spins inverse design software \cite{Su2019}. 
For the even mode dispersion engineering, we used an in-house, GPU-accelerated FDFD solver, Maxwell-B, which was further modified to perform eigen-value calculations. 
For the odd mode and the radiation loss optimization we used Legume, a guided mode expansion software capable of calculating imaginary frequencies \cite{Minkov2020}. 
A wrapper around Legume was developed by our team to make this solver compatible with the Spins software.
FDTD simulations were performed with Lumerical FDTD. 
Note, differences in the way the two simulators discretize on the Yee grid results in a small blue shift when comparing the Lumerical results with the Maxwell solver. \\

\noindent\textbf{Fabrication} The photonic crystal devices were fabricated on silicon-on-insulator with \SI{220}{\nano\meter} silicon device layer and \SI{2}{\micro\meter} of buried oxide. 
The device patterns were written in ZEP520A in a JEOL JBX-6300FS electron-beam lithography system.
After development, the pattern was etched in the silicon layer with a HBr/O\textsubscript{2}/He plasma etch. 
The resist was subsequently removed with an overnight solvent bath and a piranha clean. 
The buried oxide under the PCWs was removed with a buffered oxide etch (BOE), but was left under the grating couplers and strip waveguides. 
Microposit S1822 resist was used to protect the sample from the etch, and openings above the PCWs were patterned in the resist layer using a Durham Magneto Optics ML3 MicroWriter and a development step. 
After the BOE etch, the resist was removed with an overnight solvent bath and a piranha etch. \\

\noindent\textbf{Transmission measurements} 
Supercontinuum light from a Fianium SC400-4 was coupled into a single mode SMF-28 fiber, which was sent through an in-line fiber optic polarization controller (Thorlabs CPC250), and connected to a stripped and cleaved SMF-28 fiber\cite{Sapra2019}. 
The cleaved fiber was positioned above the input grating coupler using a holder set at \ang{5} (from vertical) inclination on a 3-axis piezo stage (Thorlabs NanoMax\texttrademark  Flexure Stages).
Similarly, a cleaved fiber was positioned above the output grating coupler using a holder at \ang{5} inclination on a 3-axis piezo stage. 
This output fiber connects to an optical spectral analyser.
To isolate the transmission spectra of the PCW device from the setup and grating coupler losses, we normalized the device measurement to a simple strip waveguide connected to an in- and output grating coupler. \\

\noindent\textbf{Dispersion measurements} 
The dispersion measurements were performed on a similar grating coupler setup as for the transmission setup. 
The cleaved fibers were replaced by OZ Optics lensed fibers (TSMJ-3A-1550-9/125-0.25-7-2.5-14-2-AR) and the holders were adjusted to a \ang{0} inclination.
Instead of the super continuum source, a tunable laser was used, and instead of the OSA we used a detector (Thorlabs DET01CFC).
For the measurements around 1300nm, Santec TSL-510 laser was used in combination with the detector connected to a oscilloscope (Agilent Technologies DSO5054A).
For the experiments around 1550nm, the Toptica CLT tunable laser was used in combination with the detector, which was connected to the signal analyser of the Toptica Digital Laser Controller. \\

\noindent\textbf{Back focal plane measurements} Detailed description of the back focal plane can be found in the Supplementary Information. Light from a Toptica CLT tunable laser is coupled to a OZ Optics lensed fiber (TPMJ-3A-1300-7/125-0.25-10-2.5-14-1-AR) positioned in a horizontal fiber holder (Thorlabs  Fiber Rotator HFR007) on a piezo stage (Thorlabs NanoMax\texttrademark Flexure Stages) and used to edge couple light into the sample. A back focal plane microscope is positioned above the sample constructed with a 60X/0.9NA Olympus Plan Fluorite objective and a Alpha NIR InGaAs-camera (Indigo Systems). An aperture in the image plane of the back focal plane microscope is used to select only light emitted by the OPA.\\

\noindent\textbf{Data availability} The data that support the plots within this paper and other findings of this study are available from the corresponding author upon reasonable request.

\medskip

\noindent\textbf{Acknowledgment}
\noindent This work is funded by the NSF ERFI NewLaw program (Award No. 1741660), and the AFOSR under the MURI program (Award No. FA9550-17-1-0002). The silicon devices were fabricated in the Stanford Nanofabrication Facility and the Stanford Nano Shared Facilities. We also thank Google for providing computational resources on the Google Cloud Platform.  \\

\noindent\textbf{Competing interests.}
The authors declare they have no competing financial interests.

\bibliography{main}

\clearpage

\end{document}